\documentclass[10pt,twocolumn,amsmath,amssymb,aps,prb,superscriptaddress]{revtex4-2}

\usepackage[pdftex]{graphicx}
\usepackage{color}
\usepackage[breaklinks=true,colorlinks,citecolor=blue,linkcolor=blue,urlcolor=blue]{hyperref}
\usepackage{graphicx}
\usepackage{natbib}
\usepackage{dcolumn}
\usepackage{bm}
\usepackage{hyperref}

\begin{document}

\title{Heitler effect and resonance fluorescence in quantum magnonics} 
\author{Enes Ilbu\u{g}a}
\affiliation{Kavli Institute of Nanoscience, Delft University of Technology, Lorentzweg 1, 2628 CJ Delft, The Netherlands}
\author{V. V. Dobrovitski}
\affiliation{Kavli Institute of Nanoscience, Delft University of Technology, Lorentzweg 1, 2628 CJ Delft, The Netherlands}
\affiliation{QuTech, Delft University of Technology, Lorentzweg 1, 2628 CJ Delft, The Netherlands}
\author{Ya. M. Blanter}
\affiliation{Kavli Institute of Nanoscience, Delft University of Technology, Lorentzweg 1, 2628 CJ Delft, The Netherlands}
%
%\affiliation{%
% Authors' institution and/or address\\
%}

%\affiliation{%
% Authors' institution and/or address\\
%}%

\date{\today}% It is always \today, today,
             %  but any date may be explicitly specified

\begin{abstract}
We consider a coupled system of a qubit and a magnon mode in which the qubit is weakly driven. We demonstrate that the spectral steady-state responses of both the qubit and the magnon show, in addition to two sidebands split by the coupling, also a peak at the driving frequency with virtually zero linewidth. This phenomenon, which persists at both strong and weak coupling, is an analog of the Heitler effect in atomic physics, and shows the path towards building of coherent magnon sources. 

\end{abstract}

%\keywords{Suggested keywords}%Use showkeys class option if keyword
                              %display desired
\maketitle

In recent years, magnonics, which studies the properties of magnons --- elementary excitations of magnetic structure, --- has been realized as a potential powerful tool in classical and quantum information transfer \cite{MagnonicsRoadmap}. To make it functional, individual magnons need to be controllably excited, manipulated, and detected. Currently, the most common instrument in manipulating magnons is cavity magnonics \cite{Yuan,ZareRameshti}. A magnet is put in a microwave or an optical cavity and interacts with the cavity field via Zeeman and magnetooptical effects. It was first theoretically proposed \cite{Flatte} and then experimentally demonstrated \cite{Hubl,Nakamura_classical} that the Zeeman effect leads to a strong coupling of a magnet to the magnetic field of a microwave cavity. This strong coupling became a necessary condition to explore quantum properties of magnons \cite{Nakamura_quantum1} with a perspective of using them for quantum information processing. For example, one of the long-standing problems is quantum state transduction between microwave and optical ranges~\cite{Lambert,Lauk,Awshalom,Matsko,Taylor,Winger}, and a recent proposal to use magnons as a part of two-stage transduction~\cite{Engelhardt} requires strong and tunable coupling.

To probe quantum magnons, the experiments add a superconducting qubit to a microwave cavity. In this setup, the role of the cavity is to couple the magnet and the qubit, and strong coupling between the two has been achieved \cite{Nakamura_quantum1,Nakamura_quantum2,Nakamura_quantum3,Nakamura_quantum4,Wolski}. This scheme is an extension of the circuit QED \cite{circuit_QED} which is the mainstream architectue for quantum circuits using superconducting qubits. Quantum magnonics is a rapidly developing field, with non-trivial quantum magnonic states already having been demonstrated \cite{You}.

There is, however, a significant obstacle towards broad application of magnons in quantum technology. The material of choice is yittirum iron garnet (YIG). This ferrimagnetic insulator has currently the highest available magnetic quality, with the quality factor which can exceed $10^4$. At the GHz frequencies, where superconducting qubits operate, this means that the lifetime of a magnon is of the order of 100 ns, much below those of a qubit and a cavity. It was theoretically proposed earlier \cite{Kounalakis,Kounalakis1,Dols} that by coupling a magnon directly to a qubit and thus departing from the cavity QED architecture one can facilitate creation of quantum states, but the magnon lifetime still would remain the limiting factor. The quality factor of YIG is a material parameter, and currently there are no prospects of increasing it by orders of magnitude \cite{ZareRameshti}.

%Heitler effect

In this work we analyze the possibility to create a long-lived stream of magnons with virtually zero linewidth by implementing a magnonic analog of resonance  fluorescence in the Heitler regime \cite{heitler1984quantum,HeAtature,MathiessenAtature1,MathiessenAtature2}. 
This regime was originally discovered and studied for an optical two-level emitter (e.g.\ atom), weakly driven by a highly monochromatic resonant electromagnetic field (e.g.\ laser). In such a situation, the incident photons from the driving beam are elastically scattered by the emitter, retaining the frequency and the phase of the driving field. As a result, a stream of indistinguishable single photons is produced, whose frequency and phase are fixed by the driving, such that the linewidth of this component of the emitted radiation is limited only by the degree of monochromaticity of the drive, not by the natural linewidth of the emitter. Such photons have numerous applications in quantum science and technology, for instance, as a valuable resource for quantum communication and entanglement generation.

In this manuscript, we analyze a similar phenomenon for a qubit coupled to a magnonic mode. Unlike the original Heitler's setting \cite{heitler1984quantum}, where the emitter, besides driving, is weakly coupled to a continuum of electromagnetic modes, we consider a typical magnonic system, where the qubit has finite lifetime $\gamma^{-1}$ and is strongly coupled to a single magnon mode, also having finite lifetime $\kappa^{-1}$.
We demonstrate that if one of the systems is weakly driven, a narrow spectral peak develops in both systems at the frequency of the driving field, with the linewidth limited only by the stability of the driving field. Thus, a stream of magnons with firmly fixed phase and frequency is produced, and highly stable long-lived  correlations between the states of the qubit and of the magnonic mode are established in a stationary state. These correlations can be used for a variety of quantum information tasks, such as high-precision sensing and metrology, or for entangling different qubits \cite{BeukersPRXQ24,ChildressPRA05,MathiessenAtature1}. 
%
%The model and the weak driving

%We consider a situation where the qubit strongly coupled to a single magnon 
%mode and driven by a monochromatic field of frequency $\omega_L$. 

We assume that the frequencies of the qubit $\omega_q$ and the magnon $\omega_m$ are close to each other, and that the coupling is linear. The linear coupling can be realized in both scenarios, when it is mediated by a cavity \cite{ZareRameshti} (in which case we consider the cavity to be integrated out) and when it is direct without a cavity \cite{Kounalakis}. The qubit is described by the Pauli matrices and the magnon mode is considered to be a harmonic oscillator. It is convenient to describe the system in the frame rotating with the frequency of the drive $\omega_L$; this is done by applying a unitary transformation to the original Hamiltonian $\hat{H}'$ and obtaining the transformed Hamiltonian $\hat{H} = \hat{D} \hat{H}' \hat{D}^{\dagger}+i \dot{\hat{D}}\hat{D}^{\dagger}$, where the operator $\hat{D} = \exp{\left[i\omega_Lt \left( \hat{\sigma}_z/2 + \hat{m}^{\dagger}\hat{m}\right)\right]}$, see Supplemental Material (SM) \cite{suppmat}. The secular part of the Hamiltonian in the rotating frame becomes 
\begin{eqnarray} 
\label{eq:hamil}
\hat{H} &=& \frac{1}{2}\Delta_q \hat{\sigma}_z + \Delta_m \hat{m}^{\dagger}\hat{m} + g\left(\hat{\sigma}_+\hat{m} + \hat{\sigma}_-\hat{m}^{\dagger}\right) \nonumber \\
&+& \Omega\left(\hat{\sigma}_+ + \hat{\sigma}_{-} \right).       
\end{eqnarray}
Here  $\hat{\sigma}_z$, $\hat{\sigma}_+$, and $\hat{\sigma}_-$ are the Pauli spin operators acting on the qubit, $\hat{m}^{\dagger}$ and $\hat{m}$ are respectively the bosonic magnon creation and annihilation operators, $\Delta_q = \omega_q - \omega_L$ and $\Delta_m = \omega_m - \omega_L$ are the detuning of qubit and magnon frequencies from the driving frequency $\omega_L$, $g$ is the interaction strength, and $\Omega$ is the driving strength. The first three terms of the Hamiltonian represent the Jaynes-Cummings model (see Ref. \cite{circuit_QED}) whereas the last term describes the external drive. For simplicity, the Planck's constant $\hbar$ is set to $1$ in this Hamiltonian and subsequent equations. 

The dissipation is included using the Markovian master equation in Lindblad representation. The master equation reads
\begin{equation}
\begin{split}\label{eq:masterequation}
\frac{d}{d t} \hat{\rho}(t) &= i\left[\hat{\rho}(t), \hat{H}\right] \\
 &+\frac{\gamma}{2}\left( 2\hat{\sigma}_- \hat{\rho}(t) \hat{\sigma}_+ -\left\{\hat{\sigma}_+ \hat{\sigma}_-, \hat{\rho}(t)\right\}\right)\\
 &+\frac{\kappa}{2}\left( 2\hat{m} \hat{\rho}(t) \hat{m}^{\dagger}-\left\{\hat{m}^{\dagger} \hat{m}, \hat{\rho}(t)\right\}\right), 
\end{split}
\end{equation}
where $\hat{\rho}$ is the density operator in the interaction picture, the square (curly) brackets denote the (anti) commutator, and $\gamma$ and $\kappa$ represent the spontaneous emission rates of the qubit and magnon, respectively. Thermally-induced emission and absorption can be disregarded at typical sub-Kelvin temperatures the system operates.

We are interested in analyzing the spectral steady-state response of the qubit and of the magnonic mode, given by \cite{scully1997quantum}
\begin{subequations}
\label{eq:spectralfs}
\begin{align}
\begin{split}
        S_q(\omega) = 2\operatorname{Re} \int_{0}^{\infty} d \tau e^{-i (\omega-\omega_L) \tau}\left\langle\hat{\sigma}_+(\tau) \hat{\sigma}_-(0)\right\rangle_{\text{ss}}\label{eq:specfuncQub}
\end{split}
\\
\begin{split}
    S_m(\omega) = 2\operatorname{Re} \int_{0}^{\infty} d \tau e^{-i (\omega-\omega_L) \tau}\left\langle\hat{m}^{\dagger}(\tau) \hat{m}(0)\right\rangle_{\text{ss}}\label{eq:specfuncMag},
\end{split} 
\end{align}
\end{subequations}
i.e.\ by the Fourier transform of the steady-state correlation functions of the qubit ${\hat\sigma_-}(t)$ and the magnon mode ${\hat m}(t)$ operators. The spectra of the correlation functions determine the dynamics of the decay and the response to driving. 

A typical form of the spectral response functions for the qubit and the magnon is shown in Fig.~1 for the case where the driving is exactly in resonance with both the qubit and the magnon, $\omega_L=\omega_q=\omega_m=\omega_0$. The two side peaks in both $S_q(\omega)$ and $S_m(\omega)$, which arise at the frequencies $\omega_0\pm g$, correspond to an expected response of two strongly coupled subsystems, with finite peak widths controlled by the  respective finite lifetimes $\gamma^{-1}$ and $\kappa^{-1}$ of the qubit and the magnon. In addition, a central $\delta$-function (zero-width) peak develops at the driving frequency $\omega_0$; this peak is a magnonic analog of the Heitler effect, and corresponds to elastic scattering of the driving field by the strongly coupled qubit-magnon system. 
This $\delta$-like peak is primarily of interest for us, since its width is limited by the very narrow bandwidth of the driving source and is independent of the decay rates of the qubit or the magnon.

\begin{figure}[!t]
\centering
\includegraphics[width=\columnwidth]{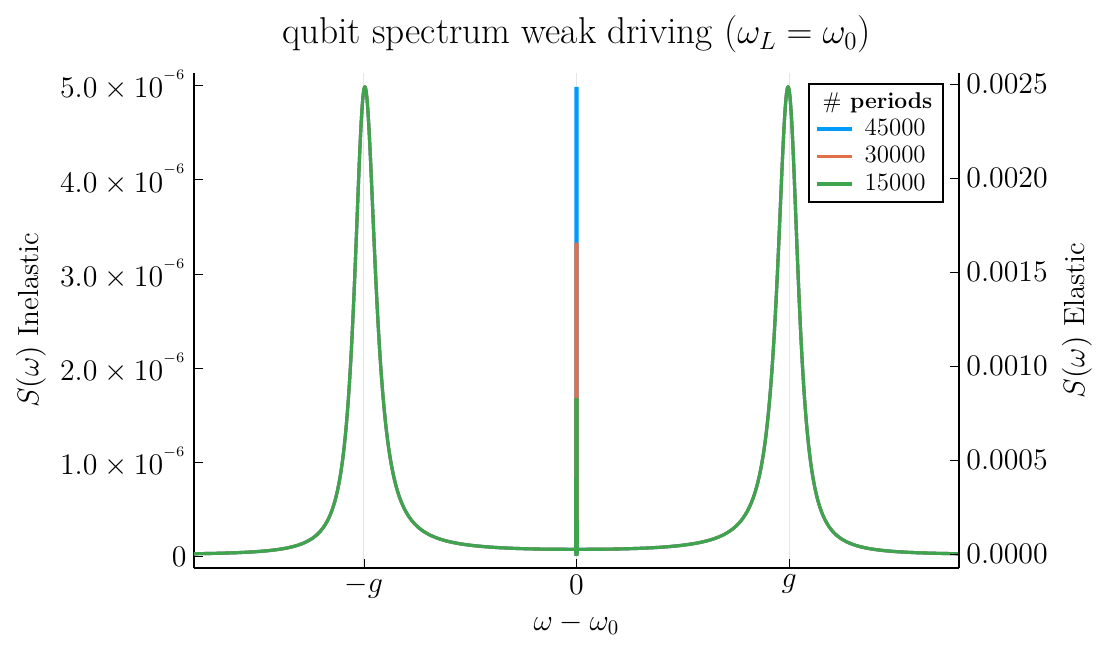}
\includegraphics[width=\columnwidth]{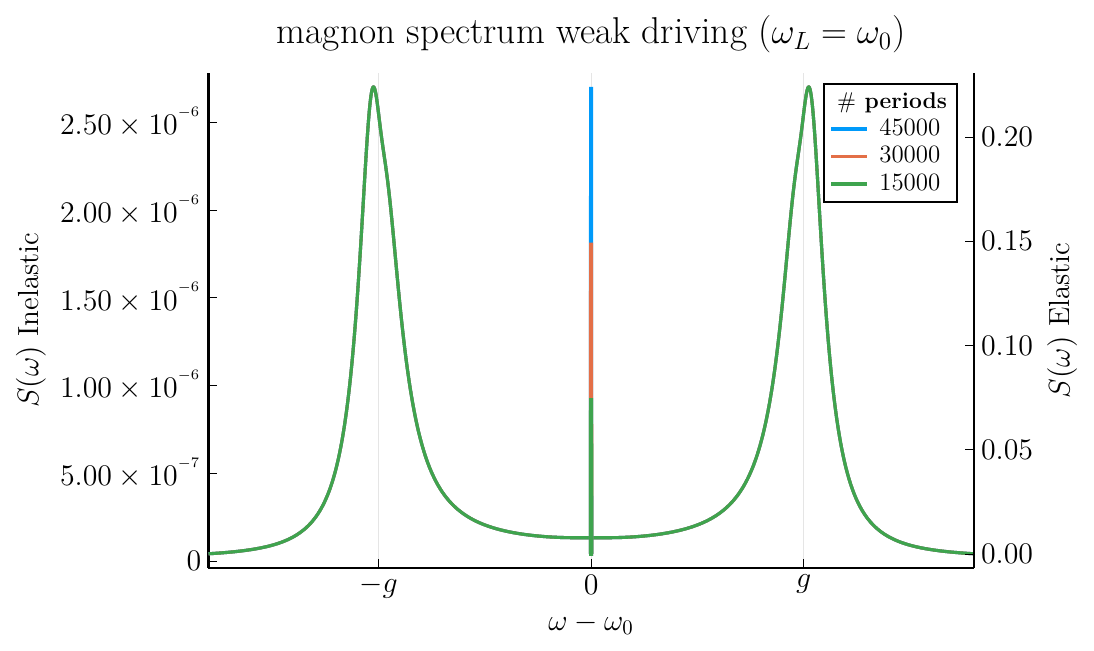}
\caption{The qubit (top) and the magnon (bottom) spectrum with a weak external drive. It is plotted for different number of time periods upon which the integral is calculated. The parameters are $g=50$, $\gamma=5$ and, $\kappa = 10$, and $\Omega=1$ (in arbitrary units)}.
\label{fig:numspec}
\end{figure}

Finding the explicit form for the correlation functions in Eq.~(\ref{eq:spectralfs}) requires solution of the master equation (\ref{eq:masterequation}), which, in general, is a daunting task. However, we are mostly interested in the weak driving regime, where the $\delta$-function peaks are most prominent \cite{cohen1998atom}. Accordingly, 
we can limit ourselves to the subspace consisting of the three states $|0\rangle \equiv|g, 0\rangle$, $|m\rangle \equiv|g, 1\rangle$, and  $|q\rangle \equiv|e, 0\rangle$, where $g$ and $e$ are the ground and the excited states of the qubit, and $0$ and $1$ stand for the presence of zero and one magnon, respectively. Solution confined to this three-state basis is valid as long as the driving amplitude satisfies $\Omega \ll \gamma$ or $\Omega \ll \kappa$. Writing the master equation in this basis gives us the following set of differential equations,
\begin{subequations}
\label{eq:diffequations}
\begin{align}
\begin{split}\label{eq:dmdt}
    \frac{d}{d t} \langle \hat{m} \rangle = &-\left(i\Delta + \kappa/2\right)\langle \hat{m} \rangle - ig\langle \hat{\sigma}_- \rangle \\ 
    &+i\Omega\langle \hat{\sigma}_+\hat{m} \rangle \ ,
\end{split}
\\
\begin{split}
    \frac{d}{d t} \langle \hat{\sigma}_- \rangle
= &-\left(i\Delta + \gamma/2\right)\langle \hat{\sigma}_- \rangle - ig\langle \hat{m} \rangle \\ 
&+ i\Omega \left(2\langle \hat{\sigma}_+\hat{\sigma}_- \rangle + \langle \hat{m}^{\dagger}\hat{m} \rangle  -1\right) \ ,
\end{split}
\\
\begin{split}
\frac{d}{d t} \langle \hat{\sigma}_+\hat{m} \rangle = &-\frac{\gamma+\kappa}{2}\langle \hat{\sigma}_+\hat{m} \rangle + ig\left(\langle \hat{m}^{\dagger}\hat{m} \rangle - \langle \hat{\sigma}_+\hat{\sigma}_- \rangle\right)  \\
&+i\Omega\langle \hat{m} \rangle \ ,
\end{split}
\\
\begin{split}
\frac{d}{d t} \langle \hat{m}^{\dagger}\hat{m} \rangle = &-\kappa\langle \hat{m}^{\dagger}\hat{m} \rangle + ig\left(\langle \hat{\sigma}_+\hat{m} \rangle - \langle \hat{\sigma}_+\hat{m} \rangle^{\dagger}\right) \ ,
\end{split}
\\
\begin{split}\label{eq:dtdsigmasigma}
\frac{d}{d t} \langle \hat{\sigma}_+\hat{\sigma}_- \rangle = &-\gamma\langle \hat{\sigma}_+\hat{\sigma}_- \rangle + ig\left(\langle \hat{\sigma}_+\hat{m} \rangle^{\dagger} - \langle \hat{\sigma}_+\hat{m} \rangle\right) 
\\ 
&+ i\Omega \left(\langle \hat{\sigma}_- \rangle - \langle \hat{\sigma}_- \rangle^{\dagger}\right) \ .
\end{split}
\end{align}
\end{subequations}
We now consider numerical and analytical solutions of these equations.

%Solutions - Numerical integration

We first integrate Eqs. (\ref{eq:dmdt}) --  (\ref{eq:dtdsigmasigma}) numerically to analyze the effect of driving for various parameter values. Then, using the discrete Fourier transform, as defined in Ref. \cite{newman2013computational}, the spectral distributions (Eqs. (\ref{eq:specfuncQub}) and (\ref{eq:specfuncMag})) are calculated. In Fig.~\ref{fig:numspec} the spectral distributions are plotted at exact resonance $\left(\omega_q = \omega_m = \omega_L \equiv \omega_0\right)$ in the strong coupling regime $\left(g \gg \kappa, \gamma\right)$.

As we discussed above, a $\delta$-function-like peak occurs at the driving frequency $\omega_0$ for both the qubit and the magnon, along with two Lorentzian sidebands at $\omega=\omega_0 \pm g$. These sidebands have the same line shape as the decay spectra of the qubit and magnon (see SM \cite{suppmat}). The central peak occurring in both the qubit and magnon spectra is not present in the decay spectra, being the result of driving the qubit. Since we make use of the discrete Fourier transform, the central peak has finite height that scales linearly with the number of time intervals $N_t$ as the total integration time grows, as seen in Fig.~\ref{fig:numspec}. Thus, the central peak indeed is not Lorentzian, but has a $\delta$-function character. It is important to note that this central peak appears only when the qubit is driven.

The stationary non-zero values of the correlation functions demonstrate that there are strong long-lived correlations between the states of the qubit and of the magnonic mode, i.e.\ beween the states $|m\rangle \equiv|g, 1\rangle$, and  $|q\rangle \equiv|e, 0\rangle$: the mutual phase between these states is locked in the stationary regime and is preserved for the whole duration of driving. Such a state constitutes a useful resource for a variety of quantum information tasks \cite{BeukersPRXQ24,ChildressPRA05,MathiessenAtature1}.

%Analytical solutions

To find the magnitude and characteristics of the peaks, we solve the system of equations analytically in the limit of weak driving. The steady-state solutions for the single time operators $\langle \hat{\sigma}_+ \rangle$ and $\langle \hat{m}^{\dagger}\rangle$ provide us with the magnitudes of the $\delta$-functions~\cite{cohen1998atom}. These steady-state solutions are readily found by rewriting the differential equations as a matrix equation (See SM \cite{suppmat}). In the weak driving limit they are
\begin{eqnarray} \label{eq:sigmassboss}
 & &  \left|\langle \hat{\sigma}_+ \rangle_{ss}\right|^2 = \\
 & & \frac{\Omega^2\left( \Delta^2 + \frac{1}{4} \kappa^2\right)}{g^4 + \Delta^4 + \frac{1}{4} \gamma^2 \Delta^2 + \frac{1}{4} \kappa^2 \Delta^2 -2 \Delta^2 g^{2}+ \frac{1}{16} \kappa^2 \gamma^2 + \frac{1}{2} g^{2} \gamma \kappa} \nonumber
\end{eqnarray}
and
\begin{eqnarray} \label{eq:magssboss}
& & \left|\langle \hat{m}^{\dagger}\rangle_{ss}\right|^2 =  \\
& &  \frac{\Omega^{2} g^{2}}{g^{4} + \Delta^{4} + \frac{1}{4} \gamma^{2} \Delta^{2} + \frac{1}{4} \kappa^{2} \Delta^{2} -2 \Delta^{2} g^{2}+ \frac{1}{16} \kappa^2 \gamma^2 + \frac{1}{2} g^{2} \gamma \kappa }, \nonumber
%\label{eq:mainresults}
\end{eqnarray}
where $\Delta = \omega_0 - \omega_L$ and $\omega_q = \omega_m \equiv \omega_0$, i.e., the qubit and magnon operate at the same frequency. Eqs.~(\ref{eq:sigmassboss}) and~(\ref{eq:magssboss}) give us the magnitudes of the $\delta$-functions. These magnitudes will now be considered in different limits.

\begin{figure}[h!]
    \includegraphics[width=0.45\textwidth]{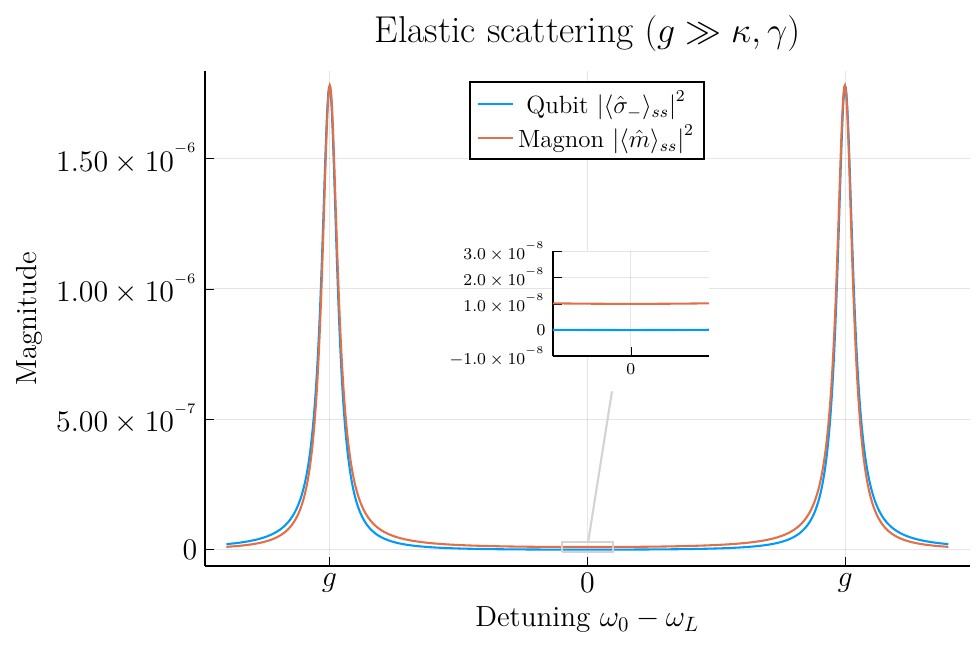}
    \caption{The elastic scattering amplitude of the qubit and the magnon plotted against the driving frequency in the strong coupling regime. The parameters are $g=100$, $\Omega = 0.01$, $\gamma = 15$ and
$\kappa = 0$ (in arbitrary units).}
    \label{fig:elasscatstrong}
\end{figure}
In the {\em strong coupling limit} ($g \gg \gamma, \kappa$), Fig.~\ref{fig:elasscatstrong} presents a plot of the magnitude of the delta-peak in the corresponding spectra for various driving frequencies. 
We note that the qubit and magnon have almost identical scattering amplitudes. The presence of the magnon magnitude in Fig.~\ref{fig:elasscatstrong} verifies that driving the qubit induces elastic scattering of the magnon.
The shapes of both magnitudes follow the shape of the decay spectrum (see SM), which is expected in the weak driving regime~\cite{heitler1984quantum}.

\begin{figure}[h!]
    \includegraphics[width=0.45\textwidth]{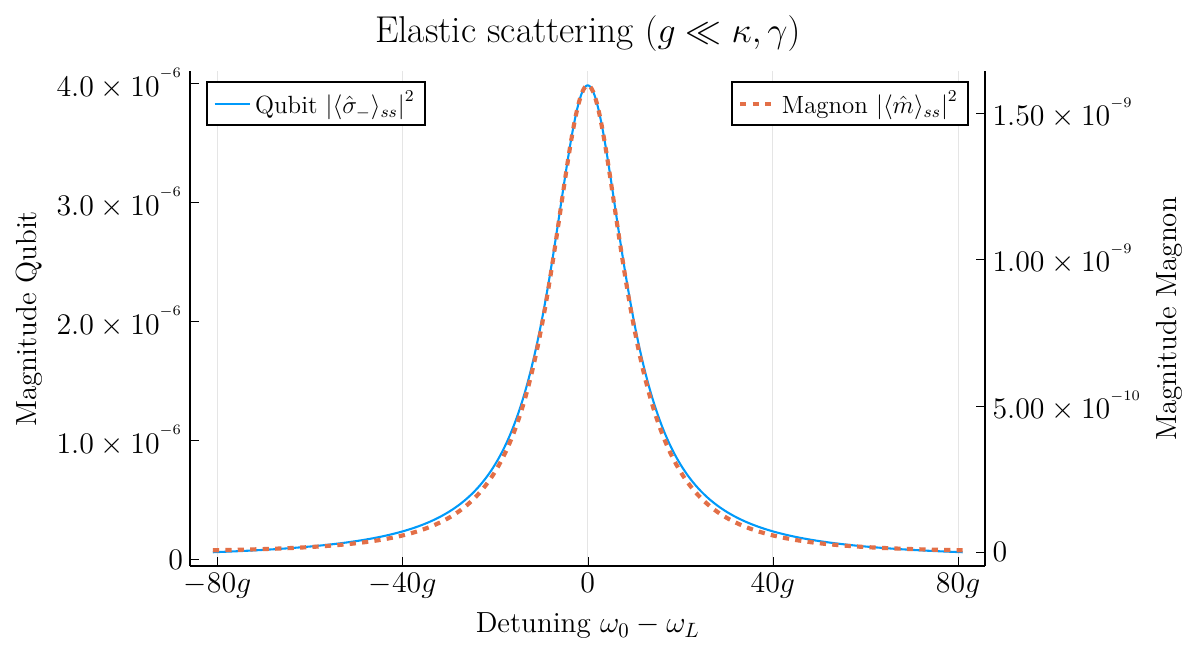}
    \caption{The elastic scattering magnitude of the qubit and the magnon plotted against the driving frequency in the weak coupling regime. The parameters are $g=0.5$, $\Omega = 0.01$, $\gamma = 10$ and
$\kappa = 50$ (in arbitrary units).}
    \label{fig:elasscatweak}
\end{figure}
For the {\em weak coupling}  ($g \ll \gamma, \kappa$), the elastic scattering magnitude is visualized in Fig. ~\ref{fig:elasscatweak}. We see that even in the weak coupling limit, the magnon emission line contains a delta-peak, albeit with a much smaller magnitude than that of the qubit. In this limit, the magnon decay spectrum is independent of the qubit decay spectrum (See SM). Here however the magnon follows the lineshape of the qubit. Physically one would expect the qubit scattering magnitude to be larger since the magnon "feels" less of the driving because of the weaker coupling strength. Even when the coupling is weak the magnet still emits single magnons. The scattering amplitudes at the maximum $\omega_0 = \omega_L$  are
\begin{equation}
\left|\langle \hat{\sigma}_+ \rangle_{ss}\right|^2 = \frac{4\Omega^2}{\gamma^2}, \ \ \ \left|\langle \hat{m}^{\dagger}\rangle_{ss}\right|^2 = \frac{16\Omega^2 g^2}{\kappa^2\gamma^2} \ .
\end{equation}
We see that the qubit scattering magnitude is independent of the decay of the magnon. This is expected since we drive the qubit. At the same time, the magnon peak grows with increasing coupling strength. The magnon peak at the maximum is smaller by a factor of $\kappa^2/(4g^2)$.
If the driving is slightly detuned from the resonance frequency, but remains within the natural linewidth, then the amplitude of the $\delta$-peak is reduced, following the natural shape of the resonance line. 

In conclusion, we have considered the system of coupled qubit and a magnon mode and found that at weak driving the spectral responses of both the qubit and the magnon develop a narrow delta-peak at the driving frequency. 
This effect is a magnonic analog of the Heitler regime of resonance fluorescence \cite{heitler1984quantum}, and represents elastic scattering of the driving by the qubit (more precisely, by the strongly coupled system of the qubit and the magnonic mode). As a result, a stationary stream of magnons is produced, whose phase and frequency are determined by the driving, with virtually zero linewidth, not limited by the qubit or the magnon lifetimes. At the same time, stable long-lived correlations between the states of the qubit and of the magnonic mode are established, with the fixed phases between the states $|0\rangle \equiv |g, 0\rangle$, $|m\rangle \equiv|g, 1\rangle$ and  $|q\rangle \equiv|e, 0\rangle$.

This effect can be used for creating a highly stable source of single magnons with fixed frequency and phase, that would be a valuable resource for magnon-based quantum sensing and metrology. Besides, the highly stable correlations between the states of the qubit and the magnon mode could be used for entangling different qubits, in analogy to a broad spectrum of entanling protocols developed in quantum optics, see {\em e.g.} Refs. \cite{BeukersPRXQ24,ChildressPRA05,EnglundReview,Matsko,Tian,Tsang,Taylor,Winger}, provided that single magnons are reliably detected with reasonable efficiency. 

The work was supported by the Dutch Research Council (NWO). This work is part of the research programme NWO QuTech Physics Funding (QTECH, programme 172) with project number 16QTECH02, which is (partly) financed by NWO.

\end{document}

    %\preprint{APS/123-QED}

\title{Supplemental Material\\Heitler effect and resonance fluorescence in quantum magnonics}

\author{Enes Ilbu\u{g}a}
\affiliation{Kavli Institute of Nanoscience, Delft University of Technology, Lorentzweg 1, 2628 CJ Delft, The Netherlands}
\author{V. V. Dobrovitski}
\affiliation{Kavli Institute of Nanoscience, Delft University of Technology, Lorentzweg 1, 2628 CJ Delft, The Netherlands}
\affiliation{QuTech, Delft University of Technology, Lorentzweg 1, 2628 CJ Delft, The Netherlands}
\author{Ya. M. Blanter}
\affiliation{Kavli Institute of Nanoscience, Delft University of Technology, Lorentzweg 1, 2628 CJ Delft, The Netherlands}
%\keywords{Suggested keywords}%Use showkeys class option if keyword
                              %display desired
\maketitle

%\tableofcontents

\section{Decay spectra}\label{sec:decayspec}

Our starting point is the Hamiltonian for a qubit resonantly coupled to a single magnon mode, see e.g. Ref. \cite{ZareRameshti},
\begin{equation} \label{eq:ham_original}
\hat{H}' = \frac{1}{2}\omega_q \hat{\sigma}_z + \omega_m \hat{m}^{\dagger} \hat{m} + g \left( \hat{\sigma}_+ \hat{m} + \hat{\sigma}_{-} \hat{m}^{\dagger} \right) + \Omega \left( \hat{\sigma}_{+} e^{-i \omega_L t} + \hat{\sigma}_{-}e^{i \omega_L t} \right) ,
\end{equation}
where $\hat{\sigma}$ are the Pauli operators describing the qubit, and $\hat{m}$ is the annihilation operator for the magnon. To derive Eq. (\ref{eq:ham_original}), we linearized the Holstein-Primakoff transformation, so that the magnon mode is described as a harmonic oscillator. Furthermore, $\omega_q$ and $\omega_m$ are the frequencies of the qubit (Rabi frequency) and the magnon, $g$ is the coupling constant, and the last term in Eq. (\ref{eq:ham_original}) describes external drive with the amplitude $\Omega$ and the frequency $\omega_L$. As in the main text, we use units with $\hbar=1$.

It is convenient to move to the frame rotating with the frequency of the drive $\omega_L$; this is done by applying a unitary transformation $\hat{H} = \hat{D} \hat{H}' \hat{D}^{\dagger}+i \dot{\hat{D}}\hat{D}^{\dagger}$, where the operator $\hat{D} = \exp{\left[i\omega_Lt \left( \hat{\sigma}_z/2 + \hat{m}^{\dagger}\hat{m}\right)\right]}$. We get the transformed Hamiltonian,
\begin{equation} 
\label{eq:ham_transformed}
\hat{H} = \frac{1}{2}\Delta_q \hat{\sigma}_z + \Delta_m \hat{m}^{\dagger}\hat{m} + g\left(\hat{\sigma}_+\hat{m} + \hat{\sigma}_-\hat{m}^{\dagger}\right) + \Omega\left(\hat{\sigma}_+ + \hat{\sigma}_{-} \right) ,       
\end{equation}
with $\Delta_{q, m} = \omega_{q,m} - \omega_L$. In the following, we assume that the qubit and the magnon are exactly at resonance, $\omega_q = \omega_m = \omega_0$, and define the detuning from the driving frequency $\Delta = \omega_0 - \omega_L$.

To calculate the decay spectra, we disregard the driving and consider the qubit initially in the excited state and the magnon with a population of one. The differential equations describing this system are

\begin{subequations}
\begin{align}
\begin{split}\label{eq:decaydiffs1}
\frac{d}{d t} \langle \hat{m}^\dagger \rangle &= \left(i\Delta - \kappa/2\right)\langle \hat{m}^\dagger \rangle + ig\langle \hat{\sigma}_+ \rangle , 
\end{split}
\\
\begin{split}\label{eq:decaydiffs2}
\frac{d}{d t} \langle \hat{\sigma}_+ \rangle &= \left(i\Delta - \gamma/2\right)\langle \hat{\sigma}_+ \rangle + ig\langle \hat{m}^\dagger \rangle,
\end{split}
\end{align}
\end{subequations}
where $\gamma$ and $\kappa$ are the decay parameters of the qubit and the magnon, respectively. We get the following solutions,

\begin{equation}\label{eq:decaydiffeqgen}
    \begin{bmatrix}
    \langle \hat{m}^\dagger \rangle(t)\\
    \langle \hat{\sigma}_+ \rangle(t)
    \end{bmatrix} = A\begin{bmatrix}
    i(\kappa-\gamma) + \mu\\
    4g
    \end{bmatrix}e^{(-(\gamma + \kappa)+ i\mu) t/4} + B\begin{bmatrix}
    i(\kappa-\gamma) - \mu\\
    4g
    \end{bmatrix}e^{(-(\gamma + \kappa)- i\mu) t/4},
\end{equation}
where $A$ and $B$ depend on the initial conditions and $\mu = \sqrt{16g^2 - (\gamma-\kappa)^2}$. To calculate the spectra, we first need expressions for the correlation functions $\langle \hat{m}^{\dagger}(t)\hat{m}(0) \rangle$ and $\langle \hat{\sigma}_+(t)\hat{\sigma}_-(0) \rangle$. By the quantum regression theorem~\cite{carmichael1999statistical}, these are given by Eq.~(\ref{eq:decaydiffeqgen}) but with $\langle \hat{m}^{\dagger}(t) \rangle$ and $\langle \hat{\sigma}_+(t)\rangle$ replaced by $\langle\hat{m}^{\dagger}(t)\hat{m}(0)\rangle$ and $\langle \hat{\sigma}_+(t)\hat{\sigma}_-(0)\rangle$ respectively. Using the initial conditions $\langle \hat{m}^{\dagger}(0)\hat{m}(0) \rangle = 1$ and $\langle \hat{\sigma}_+(0)\hat{\sigma}_-(0) \rangle = 1$, we obtain the correlation functions 
\begin{subequations}
\begin{align}
\begin{split}
        \langle \hat{m}^{\dagger}(t)\hat{m}(0) \rangle = \frac{i(\gamma - \kappa) + \mu}{2\mu}e^{(-(\gamma + \kappa)- i\mu+i\omega_0) t/4} + \frac{i(\kappa - \gamma) + \mu}{2\mu}e^{(-(\gamma + \kappa)+ i\mu+i\omega_0) t/4},
\end{split}
\\
\begin{split}
    \langle \hat{\sigma}_+(t)\hat{\sigma}_-(0) \rangle = \frac{i(\kappa - \gamma) + \mu}{2\mu}e^{(-(\gamma + \kappa)- i\mu+i\omega_0) t/4} + \frac{i(\gamma - \kappa) + \mu}{2\mu}e^{(-(\gamma + \kappa)+ i\mu+i\omega_0) t/4}.
\end{split} 
\end{align}
\end{subequations}

Calculating the spectral distribution functions~\cite{scully1997quantum} 
\begin{subequations}
\begin{align}
\begin{split}
S_m(\omega) = 2\operatorname{Re} \int_{0}^{\infty} d t e^{-i (\omega-\omega_L) t}\left\langle\hat{m}^{\dagger}(t) \hat{m}(0)\right\rangle\label{eq:specfuncMag}
\end{split}
\\
\begin{split}
    S_q(\omega) = 2\operatorname{Re} \int_{0}^{\infty} d t e^{-i (\omega-\omega_L) t}\left\langle\hat{\sigma}_+(t) \hat{\sigma}_-(0)\right\rangle\label{eq:specfuncQub},
\end{split} 
\end{align}
\end{subequations}

we find the decay spectra,
\begin{subequations}
\begin{align}
\begin{split}\label{eq:decayspecmag}
        S_{m}(\omega) &= 2\operatorname{Re}\left[\frac{i(\gamma - \kappa) + \mu}{2\mu}\frac{1}{\frac{\gamma+\kappa}{4} + i(\omega -\omega_0 + \mu/4)} + \frac{i(\kappa - \gamma) + \mu}{2\mu}\frac{1}{\frac{\gamma+\kappa}{4} + i(\omega -\omega_0 - \mu/4)}\right]
\end{split}
\\
\begin{split}\label{eq:decayspecqub}
    S_{q}(\omega) &= 2\operatorname{Re}\left[\frac{i(\kappa - \gamma) + \mu}{2\mu}\frac{1}{\frac{\gamma+\kappa}{4} + i(\omega-\omega_0  + \mu/4)} + \frac{i(\gamma- \kappa ) + \mu}{2\mu}\frac{1}{\frac{\gamma+\kappa}{4} + i(\omega-\omega_0  - \mu/4)}\right].
\end{split} 
\end{align}
\end{subequations}

These two expressions are similar, but their difference will become apparent when we consider different limits. 

\subsection{Strong coupling limit}
We start by considering the strong coupling limit, $g \gg \kappa,\gamma$. In this limit, the parameter $\mu$ can be approximated as real and equal to $\mu = 4g$. Substituting this value of $\mu$ into equations~(\ref{eq:decayspecmag}) and~(\ref{eq:decayspecqub}) gives us the following two-peak spectra,
\begin{equation}\label{eq:decayspecmagSTRon}
S_{m}(\omega) = \frac{-(\gamma - \kappa)(\omega_0-\omega) /4g+\gamma /2}{\left( \frac{\gamma+\kappa}{4}\right)^2 + \left[\omega-\left( \omega_0  - g \right)\right]^2} + \frac{-(\gamma - \kappa)(\omega-\omega_0)/4g+ \gamma /2}{\left( \frac{\gamma+\kappa}{4}\right)^2 + \left[\omega-\left( \omega_0  + g \right)\right]^2} ,
\end{equation}
\begin{equation}\label{eq:decayspecquSTRon}
S_{q}(\omega) = \frac{(\gamma - \kappa)(\omega_0-\omega)/4g+\kappa /2}{\left( \frac{\gamma+\kappa}{4}\right)^2 + \left[\omega-\left( \omega_0  - g \right)\right]^2} + \frac{(\gamma - \kappa)(\omega-\omega_0)/4g+ \kappa /2}{\left( \frac{\gamma+\kappa}{4}\right)^2 + \left[\omega-\left( \omega_0  + g \right)\right]^2}.
\end{equation}

\begin{figure}[h]
    \centering
    \includegraphics[width=0.8\textwidth]{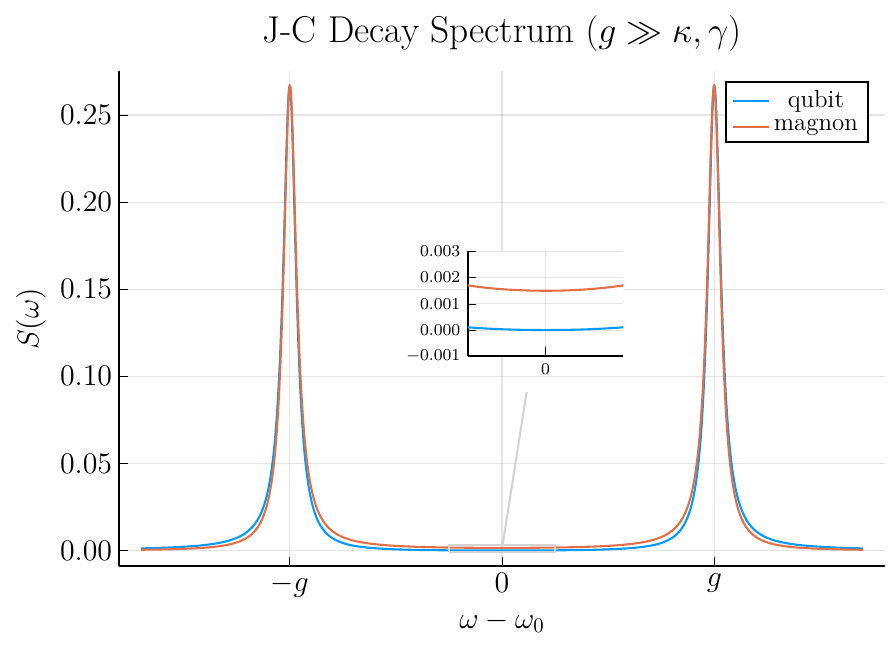}
    \caption{The decay spectrum of the qubit and the magnon within the Jaynes-Cummings model. Plotted for $g = 100$, $\gamma = 15$, $\kappa =0$.}
    \label{fig:decayspecJC}
\end{figure}

In Fig.~\ref{fig:decayspecJC},  the qubit and the magnon spectrum are plotted as a function of the frequency $\omega$.

The decay spectra show two peaks centered around the energies $\omega_0 \pm g$, which correspond to the dressed states from the Jaynes Cummings model~\cite{gerry2005introductory}. The near equivalence of both spectra is due to the fact that both systems "blend" into one system due to their strong interaction. Even without magnon decay ($\kappa = 0$) in Fig.~\ref{fig:decayspecJC}, the spectra are still nearly equivalent.

\subsection{Weak coupling limit}
In the weak coupling limit, we have $g \ll \kappa,\gamma$. We expect the qubit and magnon to become disentangled in this limit and show independent decay spectra. Taking this limit in equations~(\ref{eq:decayspecmag}) and~(\ref{eq:decayspecqub}) gives the following spectra,
\begin{equation}
    S_m(\omega) =  2\operatorname{Re}\left[\frac{1}{ \frac{\kappa}{2} + i (\omega-\omega_0)}\right]
\end{equation}
and 
\begin{equation}
    S_q(\omega) =  2\operatorname{Re}\left[\frac{1}{ \frac{\gamma}{2} + i(\omega-\omega_0)}\right].
\end{equation}

\begin{figure}[h]
    \centering
    \includegraphics[width=0.8\textwidth]{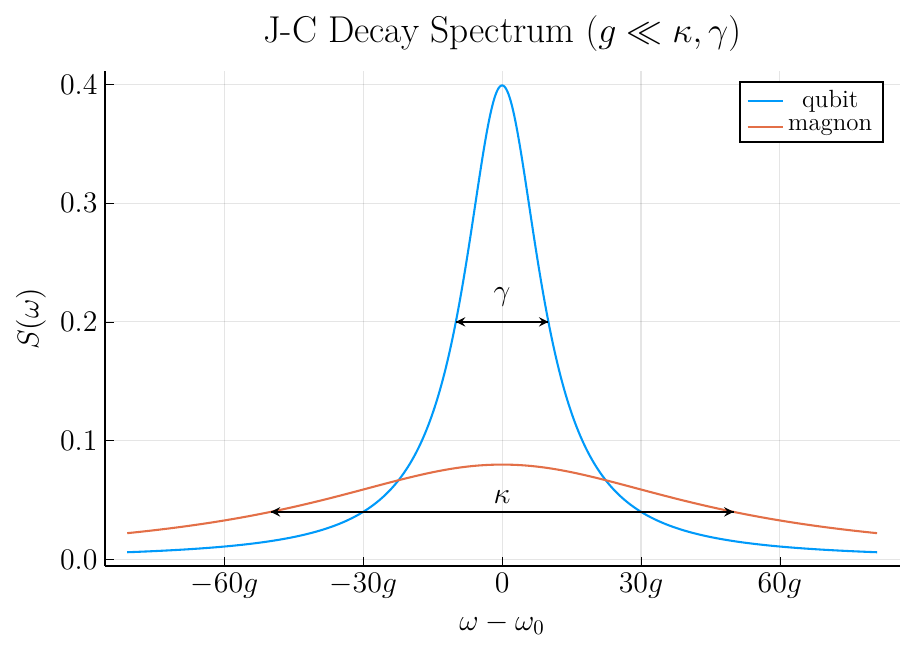}
    \caption{The decay spectrum of the qubit and the magnon within the Jaynes-Cummings model. Plotted for $g = 0.5$, $\gamma = 10$, $\kappa =50$.}
    \label{fig:decayspecweakJC}
\end{figure}

These equations indeed confirm what we would expect physically to occur. The decay spectra are once again plotted in Fig.~\ref{fig:decayspecweakJC} but now for the weak coupling regime. 

\section{Analytical solution, elastic scattering}
In comparison to the set of differential equations of section~\ref{sec:decayspec}, we consider now nonzero driving. The set of differential equations is then

\begin{subequations}
\label{eq:diffequations}
\begin{align}
\begin{split}\label{eq:dmdt}
    \frac{d}{d t} \langle \hat{m} \rangle = &-\left(i\Delta + \kappa/2\right)\langle \hat{m} \rangle - ig\langle \hat{\sigma}_- \rangle \\ 
    &+i\Omega\langle \hat{\sigma}_+\hat{m} \rangle
\end{split}
\\
\begin{split}
    \frac{d}{d t} \langle \hat{\sigma}_- \rangle
= &-\left(i\Delta + \gamma/2\right)\langle \hat{\sigma}_- \rangle - ig\langle \hat{m} \rangle \\ 
&+ i\Omega \left(2\langle \hat{\sigma}_+\hat{\sigma}_- \rangle + \langle \hat{m}^{\dagger}\hat{m} \rangle  -1\right)
\end{split}
\\
\begin{split}
\frac{d}{d t} \langle \hat{\sigma}_+\hat{m} \rangle = &-\frac{\gamma+\kappa}{2}\langle \hat{\sigma}_+\hat{m} \rangle + ig\left(\langle \hat{m}^{\dagger}\hat{m} \rangle - \langle \hat{\sigma}_+\hat{\sigma}_- \rangle\right)  \\
&+i\Omega\langle \hat{m} \rangle
\end{split}
\\
\begin{split}
\frac{d}{d t} \langle \hat{m}^{\dagger}\hat{m} \rangle = &-\kappa\langle \hat{m}^{\dagger}\hat{m} \rangle + ig\left(\langle \hat{\sigma}_+\hat{m} \rangle - \langle \hat{\sigma}_+\hat{m} \rangle^{\dagger}\right)
\end{split}
\\
\begin{split}\label{eq:dtdsigmasigma}
\frac{d}{d t} \langle \hat{\sigma}_+\hat{\sigma}_- \rangle = &-\gamma\langle \hat{\sigma}_+\hat{\sigma}_- \rangle + ig\left(\langle \hat{\sigma}_+\hat{m} \rangle^{\dagger} - \langle \hat{\sigma}_+\hat{m} \rangle\right) 
\\ 
&+ i\Omega \left(\langle \hat{\sigma}_- \rangle - \langle \hat{\sigma}_- \rangle^{\dagger}\right).
\end{split}
\end{align}
\end{subequations}

 From these equations, we see that the driving term $\Omega$ couples the  operators $\langle \hat{m}\rangle$ and $\langle\hat{\sigma}_-\rangle$ to the operators $\langle \hat{m}^{\dagger}\hat{m} \rangle$, $\langle \hat{\sigma}_+\hat{\sigma}_- \rangle$ and $\langle \hat{\sigma}_+\hat{m} \rangle$. These equations can once again be put into matrix form. To work with only real variables, we will use the real and imaginary parts of the complex operators instead of the complex variable and its conjugate. So, for example, instead of $\langle \hat{\sigma}_+\hat{m} \rangle$ and $\langle \hat{\sigma}_+\hat{m} \rangle^{\dagger}$, we use $\langle \hat{\sigma}_+\hat{m} \rangle_R$ and $\langle \hat{\sigma}_+\hat{m} \rangle_I$. The equations in matrix form are then 
\begin{equation}\label{eq:diffeqMatrixform}
   \frac{d}{d t} \mathbf{v} = \mathbf{A}\mathbf{v} + \mathbf{b},
\end{equation}
where 
\begin{equation}
    \mathbf{v} = \begin{bmatrix}
    \langle \hat{m} \rangle_R\\
    \langle \hat{m} \rangle_I\\
    \langle \hat{\sigma}_- \rangle_R\\
    \langle \hat{\sigma}_- \rangle_I\\
    \langle \hat{\sigma}_+\hat{m} \rangle_R\\
    \langle \hat{\sigma}_+\hat{m} \rangle_I\\
    \langle \hat{m}^{\dagger}\hat{m} \rangle\\
    \langle \hat{\sigma}_+\hat{\sigma}_- \rangle
    
    \end{bmatrix}
\end{equation}
,
\begin{equation}\label{eq:Fullmatrix}
\mathbf{A} = \left[
\begin{array}{cccccccc}
 - \frac{1}{2} \kappa & \Delta & 0 & g & 0 &  - \Omega & 0 & 0 \\
 - \Delta &  - \frac{1}{2} \kappa &  - g & 0 & \Omega & 0 & 0 & 0 \\
0 & g &  - \frac{1}{2} \gamma & \Delta & 0 & 0 & 0 & 0 \\
 - g & 0 &  - \Delta &  - \frac{1}{2} \gamma & 0 & 0 & \Omega & 2 \Omega \\
0 &  - \Omega & 0 & 0 & -\frac{\left(   \gamma + \kappa \right)}{2}  & 0 & 0 & 0 \\
\Omega & 0 & 0 & 0 & 0 & -\frac{\left(   \gamma + \kappa \right)}{2}  & g &  - g \\
0 & 0 & 0 & 0 & 0 &  - 2 g &  - \kappa & 0 \\
0 & 0 & 0 &  - 2 \Omega & 0 & 2 g & 0 &  - \gamma \\
\end{array}
\right]
\end{equation}
and
\begin{equation}
    \mathbf{b} = \begin{bmatrix}
    0\\
    0\\
    0\\
    -\Omega\\
    0\\
    0\\
    0\\
    0
    
    \end{bmatrix}.
\end{equation}
What we have now is a non-homogeneous system of first-order differential equations. However, the magnitudes of the $\delta$-functions are provided by the steady-state solutions of the single time operators $\langle \hat{\sigma}_- \rangle$ and $\langle \hat{m}\rangle$~\cite{cohen1998atom}. These steady-state solutions are then calculated with the formula
\begin{equation}
    \mathbf{v}_{ss} = -\mathbf{A}^{-1}\mathbf{b}.
\end{equation}
%The element  of the matrix $A^{-1}$ will be calculated through the formula
%\begin{equation}
%    \left(\mathbf{A}^{-1}\right)_{i j}=\frac{\mathbf{C}_{j i}}{\det(\mathbf{A})}
%\end{equation}
%where $\mathbf{C}$ is the matrix of cofactors. The co-factor $\mathbf{C}_{j i}$ is the product of the minor matrix $M_{ji}$ and $(-1)^{i+j}$. Finding the inverse of $\mathbf{A}$ thus consists of finding the determinant of 64 seven by seven minor matrices and the determinant of 1 eight by eight matrix. But since $\mathbf{b}$ only contains one nonzero term, the number of minor matrices of which the determinant needs to be calculated is 8. 
The elastic scattering magnitudes are given by

\begin{eqnarray}
\label{eq:sigmassboss}
   \left|\langle \hat{\sigma}_- \rangle_{ss}\right|^2 =  \frac{\Omega^{2}\left( \Delta^{2} + \frac{1}{4} \kappa^{2}\right)}{g^{4} + \Delta^{4} + \frac{1}{4} \gamma^{2} \Delta^{2} + \frac{1}{4} \kappa^{2} \Delta^{2} -2 \Delta^{2} g^{2}+ \frac{1}{16} \kappa^{2} \gamma^{2} + \frac{1}{2} g^{2} \gamma \kappa } , \\
\label{eq:magssboss}
   \left|\langle \hat{m}\rangle_{ss}\right|^2 =  \frac{\Omega^{2} g^{2}}{g^{4} + \Delta^{4} + \frac{1}{4} \gamma^{2} \Delta^{2} + \frac{1}{4} \kappa^{2} \Delta^{2} -2 \Delta^{2} g^{2}+ \frac{1}{16} \kappa^{2} \gamma^{2} + \frac{1}{2} g^{2} \gamma \kappa }.
\end{eqnarray}

% The \nocite command causes all entries in a bibliography to be printed out
% whether or not they are actually referenced in the text. This is appropriate
% for the sample file to show the different styles of references, but authors
% most likely will not want to use it.
%\nocite{*}